\documentclass{article}

    \usepackage[preprint,nonatbib]{neurips_2024}

\usepackage[utf8]{inputenc} 
\usepackage[T1]{fontenc}    
\usepackage{hyperref}       
\usepackage{url}            
\usepackage{booktabs}       
\usepackage{amsfonts}       
\usepackage{nicefrac}       
\usepackage{microtype}      
\usepackage{xcolor}         
\usepackage{graphicx}
\usepackage{bbding}

\usepackage{listings}
\lstdefinestyle{go}{
    language=Go,
    frame=lines,
    basicstyle=\ttfamily\small,
    keywordstyle=\color{blue},
    commentstyle=\color{teal},
    stringstyle=\color{red},
    numbers=left,
    numberstyle=\tiny\color{gray},
    stepnumber=1,
    numbersep=5pt,
    showstringspaces=false,
    breaklines=true,
    breakatwhitespace=true,
    tabsize=4,
    captionpos=b
}
\newcommand{\toolname}{\texttt{MarsCode Agent}~}

\title{MarsCode Agent: AI-native Automated Bug Fixing}

\author{%
  Yizhou Liu$^{1}$\thanks{Equal contribution. Order determined by rolling the dice.}\quad Pengfei Gao$^{1}$\footnotemark[1] \quad Xinchen Wang$^{2}$\thanks{Work done during an internship at ByteDance} \quad Jie Liu$^{1}$ \\
  \textbf{Yexuan Shi$^{1}$ \quad Zhao Zhang$^{1}$ \quad Chao Peng$^{1}$\thanks{Corresponding author}  } \\
  $^1$ByteDance \quad $^2$Harbin Institute of Technology, Shenzhen \\
  \texttt{\url{pengchao.x@bytedance.com}} \\
  \texttt{\url{https://se-research.bytedance.com/}} \\
}

\begin{document}

\maketitle

\begin{abstract}
Recent advances in large language models (LLMs) have shown significant potential to automate various software development tasks, including code completion, test generation, and bug fixing.
However, the application of LLMs for automated bug fixing remains challenging due to the complexity and diversity of real-world software systems.
In this paper, we introduce \toolname, a novel framework that leverages LLMs to automatically identify and repair bugs in software code.
\toolname combines the power of LLMs with advanced code analysis techniques to accurately localize faults and generate patches.
Our approach follows a systematic process of planning, bug reproduction, fault localization, candidate patch generation, and validation to ensure high-quality bug fixes.
We evaluated \toolname on SWE-bench, a comprehensive benchmark of real-world software projects, and our results show that \toolname achieves a high success rate in bug fixing compared to most of the existing automated approaches.
\end{abstract}

\section{Introduction}
The automation of software engineering tasks has long been a goal for researchers and practitioners in the field.
Recent progress in large language models (LLMs) like GPT-4o and Doubao Pro has brought us closer to this vision, enabling significant advancements in code generation, program repair, and other software development activities.
In this trend, LLM-based agents - intelligent entities capable of perceiving the external environment, operating tools, and making autonomous decisions, have garnered increasing attention from both the research and industry community.

Bug fixing is a critical aspect of software maintenance, covering tasks such as identifying the root cause of defects, generating correct patches, and validating the fixes to ensure that they do not introduce new issues.
Traditional approaches to automated bug fixing rely heavily on manually crafted rules and heuristics, which can be limited in scope and adaptability.
Recent efforts have explored the use of LLMs and LLM-based agents to address these limitations by leveraging their ability to understand and generate code in a more flexible manner.

However, applying LLMs to the automated bug fixing of real-world software projects presents unique challenges.
Unlike simple, self-contained coding tasks, bug fixing often requires an in-depth understanding of complex codebases, interdependencies among files, and context-specific issues that arise during software development.

In this report, we introduce \toolname, a novel framework designed to automate the bug fixing process using LLMs. By building an agent framework and providing interactive interfaces and tools for code retrieval, debugging, and editing, \toolname has made it possible for agents to take over some software engineering tasks.

Core contributions of \toolname include:

\begin{itemize}
    \item \toolname has developed a multi-agent collaboration framework that allocates static or dynamic solving pipelines based on the nature of the problem to be addressed, thereby flexibly adapting to various bug fixing challenges.
    \item \toolname \toolname combines code knowledge graphs and language server protocols to provide agents with comprehensive capabilities for code entity retrieval, relationship retrieval, and definition-and-reference navigation, enabling agents to browse and analyze code similarly to human developers.
    \item For code editing, \toolname uses conflict-based code edit descriptions and static syntax checking to accurately generate well-formatted code patches.
    \item In dynamic software debugging, \toolname leverages a containerized sandbox environment based on Docker, equipping agents with human developer-like debugging capabilities such as defect reproduction, log addition, and test framework execution.
\end{itemize}

These advancements underscore the potential of LLM-based AI agents in automating and enhancing various aspects of bug fixing, paving the way for more efficient and effective software engineering practices, such as feature development.

We evaluate \toolname on SWE-bench\cite{jimenez2023swe}, a diverse benchmark of real-world software projects, demonstrating its ability to effectively fix a wide range of bugs without human intervention. Our experimental results show that \toolname outperforms most of existing automated bug fixing tools in terms of both accuracy and efficiency.

By leveraging the capabilities of LLMs in a systematic and structured manner, \toolname represents a significant step forward in the quest for fully autonomous software maintenance. We believe that our framework will inspire further research and innovation in this area, ultimately leading to more robust and reliable software systems


\section{Background and Related Work}
In this section, we discuss basic concepts of large language models and their application on software engineering tasks, especially for fault localization and automated program repair.
We also talk about recent advances in LLM-based agents for software engineering and the SWE-bench benchmark for their evaluation.

\subsection{Large Language Models}

Large language models (LLMs) are highly advanced pre-trained language models.
These models undergo initial unsupervised training on vast amounts of corpus, followed by fine-tuning for specific tasks to enhance performance.
In natural language processing (NLP), LLMs have been extensively applied to various tasks such as machine translation~\cite{wang2023document, zhang2023prompting}, text summarization~\cite{zhang2023benchmarking}, and classification~\cite{mayer2023prompt}. 

Language models are classified into three categories based on their architecture: encoder-only models~\cite{feng2020codebert}, decoder-only models~\cite{nijkamp2022codegen}, and encoder-decoder models~\cite{tian2022learning}.
Most existing LLMs for code utilize the transformer architecture's encoders, known for their exceptional learning capabilities and scalability.
Regardless of their architecture, most models can be fine-tuned with task-specific data to enhance performance~\cite{lin2023cct5}.

Large language models (LLMs) have become a promising choice for various software engineering tasks due to their impressive performance in both code generation and understanding~\cite{yang2023enhancing}.
Researchers and developers have applied LLMs to several software engineering tasks, such as program synthesis~\cite{liu2024your, zhu2024sketch, wang2023practitioners, wang2023two, wang2023natural}, code translation~\cite{yu2024Mal, yang2024exploring}, program repair~\cite{lin2024one, jiang2023impact, xia2023automated}, fault detection and localization~\cite{du2024generalization, qin2024agentfl}, incident analysis~\cite{chen2024automatic, ahmed2023recommending}, code summarization~\cite{geng2024large} and testing~\cite{sun2023smt}.
For example, Codex~\cite{chen2021evaluating}, StarCoder~\cite{lozhkov2024starcoder}, and DeepSeek-Coder~\cite{zhu2024deepseek} are notable code-specific LLMs developed through extensive training on large datasets of open-source code snippets.
Additionally, instruction-following code-specific LLMs such as DeepSeek-Coder-Instruct~\cite{zhu2024deepseek} and Magicoder~\cite{wei2023magicoder} have been created using instruction-tuning methods to enhance their utility in coding tasks.

\subsection{Fault Localization}

Fault localization (FL)~\cite{wong2016survey} techniques aim to discover and analyze the location and causes of faults, which can be categorized into dynamic and static approaches.
Dynamic FL techniques, such as spectrum-based fault localization (SBFL)~\cite{abreu2007accuracy, abreu2009practical} and mutation-based fault localization (MBFL)~\cite{papadakis2015metallaxis}, analyze the dynamic execution information of a program to determine fault locations, though they are resource-intensive.
Static FL techniques~\cite{mao2014slice} determine fault locations through semantic or syntactic analysis at the bug report or source code level, offering fast detection with low resource consumption.
Advanced FL techniques, such as multiple fault localization (MFL) and combined dynamic and static methods, have emerged to guide APR tools in finding and fixing more errors~\cite{xiao2021albfl, kim2019precise, neelofar2017improving}.

\subsection{Automated Program Repair}

Automated program repair (APR)~\cite{le2019automated} has attracted significant attention over the past decade.
APR techniques aim to generate patches for buggy programs to pass given test suites.
These techniques can be categorized into search-based~\cite{li2022improving, mehne2018accelerating}, semantics-based~\cite{le2017jfix, nguyen2013semfix, le2016empirical}, and pattern/learning-based approaches~\cite{li2020dlfix, li2022dear, zhang2023survey}.
Search-based APR techniques like GenProg~\cite{le2011genprog} use predefined code mutation operators to generate patches, while semantics-based APR techniques generate patches by solving repair constraints based on test suite specifications.
Learning-based APR techniques, such as those utilizing deep learning models, train on large code repositories to predict correct patches.
Recent work has shown the use of LLMs for APR, often focusing on constructing APR-specific prompts to guide LLMs in generating patches for buggy program statements~\cite{xia2023automated}.

\subsection{Agents for Software Development}

The emergence and popularity of agent-based frameworks have led to the development of agent-based approaches for solving software engineering tasks.
Devin and its open-source counterpart OpenDevin~\cite{wang2024opendevin} are among the first end-to-end LLM agent-based frameworks.
These frameworks use agents for planning based on user requirements and enable agents to iteratively perform tasks using tools like file editors, terminals, and web search engines.
SWE-agent~\cite{yang2024swe}, for example, designs a custom agent-computer interface (ACI) allowing LLM agents to interact with the repository environment through actions such as reading, editing files, and running bash commands.
AutoCodeRover~\cite{zhang2024autocoderover} provides LLM agents with specific APIs to effectively identify locations needing modification to resolve issues. Numerous other agent-based approaches have been developed, both in open-source and commercial products.

\subsection{SWE-bench}
\label{sec:swe-bench}

SWE-Bench~\cite{jimenez2023swe} is a comprehensive benchmark designed to evaluate LLMs on complex real-world software engineering tasks sourced from GitHub issues and corresponding pull requests across 12 popular Python repositories.
This benchmark addresses the limitations of existing coding benchmarks such as HumanEval~\cite{chen2021evaluating} by presenting tasks that require models to understand and coordinate changes across large codebases involving multiple functions and files.
The benchmark includes 2,294 task instances and emphasizes the need for models to interact with execution environments and handle long contexts, showcasing the challenges that real-world software engineering problems pose to current LLMs.
Their evaluations reveal that even the best-performing models at the time of publication, such as Claude 2, achieve a success rate of only 1.96\%, highlighting significant room for improvement.

\section{Approach}
In this section, we present our approach, \toolname, spreading into the multi-agent collaborative framework, code indexing and code editing tools designed for agents.

\subsection{Multi-agent Collaborative Framework}

In our daily development work, developers often encounter various issues such as:

\begin{itemize}
    \item Test case failures, which may include errors or exception stacks due to logic errors or failed test assertions.
    \item Code output not meeting expectations, with no explicit error messages but clear expected results.
    \item The need to extend existing functionality or add new features, with clear development requirements and expected outcomes, but uncertainty about how and where to implement them.
    \item Simple defect fixes, with a rough idea of the solution but requiring assistance due to unfamiliarity with language features.
\end{itemize}

These diverse program repair and development tasks cannot be smoothly handled with a fixed approach.
For instance, some simple defect fixes or feature extensions can be completed through reviewing existing relevant code, while deeper exception stacks or complex logic errors may require dynamic code execution and variable tracking to uncover and fix the defects.

\begin{figure}
    \centering
    \includegraphics[width=0.75\linewidth]{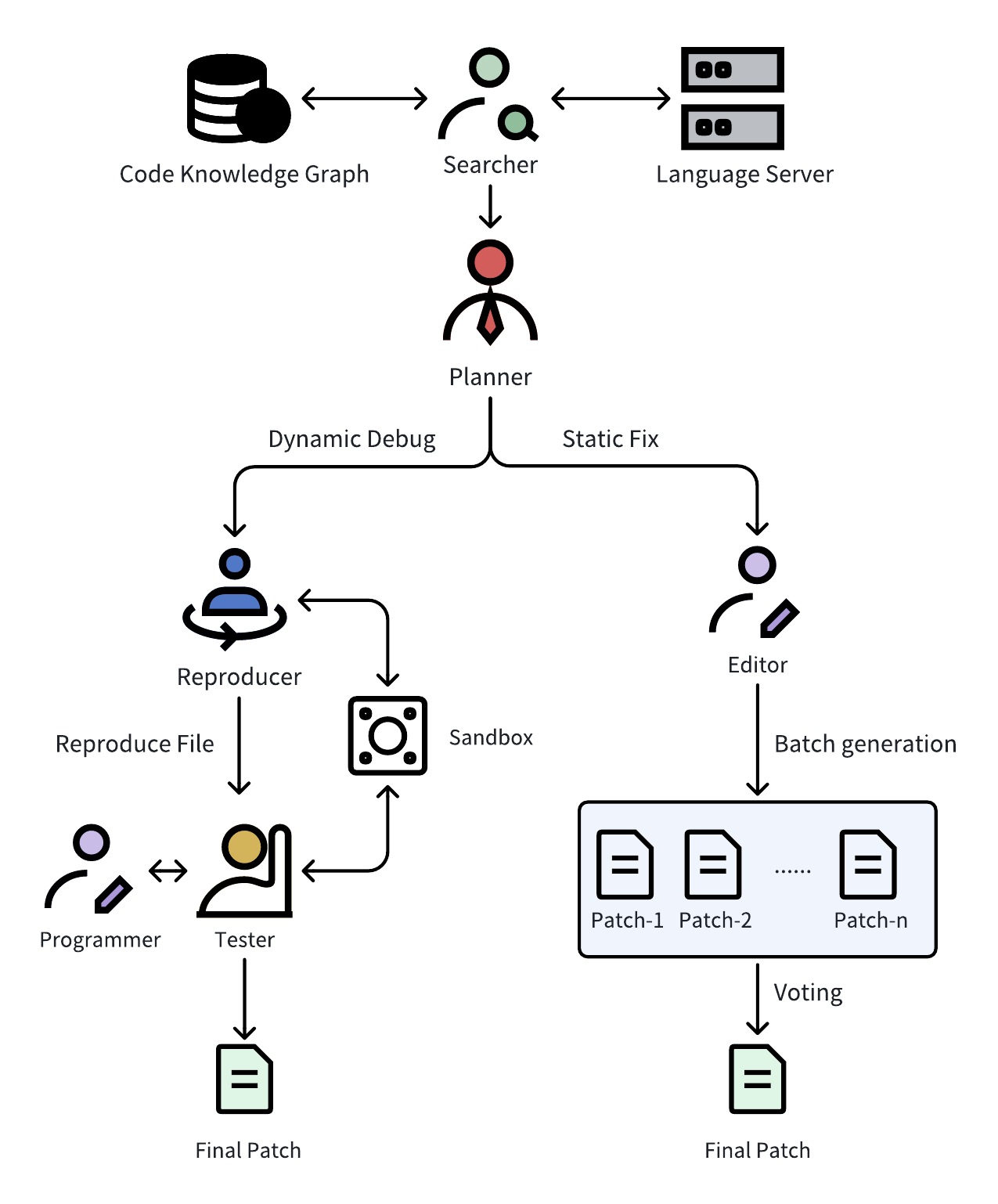}
    \caption{Multi-agent Collaborative Framework}
    \label{fig:framework}
\end{figure}

Therefore, we have adopted a multi-agent collaboration framework to adapt to different development scenarios.
As shown in Figure~\ref{fig:framework}, the framework includes the following six roles:

\begin{itemize}
    \item \texttt{Searcher} utilizes tools like code knowledge graph (CKG) and language server protocol (LSP) to collect code snippets from the repository related to the current issue.
    \item \texttt{Planner} qualitatively analyzes the collected code snippets and classifies the issue into dynamic debugging repair or static repair workflows.
    \item \texttt{Reproducer}, in dynamic debugging repair scenarios, writes reproduction scripts based on the relevant code and issue description, and performs dynamic debugging in a sandbox to confirm successful reproduction.
    \item \texttt{Programmer} edits the code according to the issue description and relevant code, iterating modifications based on the Tester’s results.
    \item \texttt{Tester} dynamically verifies the current code version using the reproduction script, checking if the issue is resolved.
    \item \texttt{Editor} attempts to provide multiple repair solutions based on the issue description and relevant code snippets, using a voting mechanism to determine the final repair result.
\end{itemize}	

We have equipped different agents with corresponding toolsets to support their tasks, as shown in Table~\ref{table:tools}. Notably, we do not grant all tools to every agent but rather limit the capabilities and responsibilities of each agent to reduce the difficulty of solving issues in each phase and improve the stability and quality of task execution.

\begin{table}[]
\caption{Tool Support of Agent Tasks}
\label{table:tools}
\centering
\scriptsize{
\begin{tabular}{lllllll}
\hline
Tool                          & \texttt{Searcher} & \texttt{Planner} & \texttt{Reproducer} & \texttt{Programmer} & \texttt{Tester} & \texttt{Editor} \\ \hline
CKG                           & \CheckmarkBold &  \CheckmarkBold       &    \CheckmarkBold        &   \CheckmarkBold         &   \CheckmarkBold     &   \XSolidBrush     \\
LSP                           & \CheckmarkBold &  \CheckmarkBold       &    \CheckmarkBold        &   \CheckmarkBold         &   \CheckmarkBold     &   \XSolidBrush     \\
General File Indexing         & \CheckmarkBold &  \CheckmarkBold       &    \CheckmarkBold        &   \CheckmarkBold         &   \CheckmarkBold     &   \XSolidBrush     \\
General Bash Command          & \CheckmarkBold &  \CheckmarkBold       &    \CheckmarkBold        &   \CheckmarkBold         &   \CheckmarkBold     &   \XSolidBrush     \\
Code Editing                  &   \XSolidBrush          &     \XSolidBrush       &   \XSolidBrush            &     \CheckmarkBold       &   \XSolidBrush     &    \CheckmarkBold    \\
Reset Repository              &   \XSolidBrush          &     \XSolidBrush       &   \XSolidBrush            &     \CheckmarkBold       &   \XSolidBrush     &   \XSolidBrush    \\
Reproduction Script Execution &    \XSolidBrush      &   \XSolidBrush      &    \CheckmarkBold          &   \XSolidBrush         &    \CheckmarkBold          &     \XSolidBrush    \\ \hline
\end{tabular}
}
\end{table}

In dynamic debugging repair scenarios, the collaboration workflow of the agents is as follows:

\begin{enumerate}
    \item The \texttt{Reproducer} creates a reproduction script matching the issue description.
    \item The reproduction script is provided to the \texttt{Tester} for verification, which then supplies the resulting exception stack and other output information to the \texttt{Programmer} for repair.
    \item After the \texttt{Programmer} completes the repair, a testing request is made to the \texttt{Tester}.
    \item The \texttt{Tester} verifies the repair using the reproduction script and determines if the issue is resolved:
        \begin{enumerate}
            \item If resolved, the diff tool is used to capture the code changes as the repair solution, ending the dynamic debugging.
            \item If unresolved, the exception stack and other output information from the reproduction process are returned to the \texttt{Programmer}.
        \end{enumerate}
    \item The \texttt{Programmer} may continue modifications based on the \texttt{Tester}’s error messages or reset the repository and start anew until the \texttt{Tester} confirms the issue is resolved.
\end{enumerate}

During this process, we set up a runtime sandbox environment in a Docker container to achieve dynamic debugging, issue reproduction, and validation.

In static repair scenarios, the agent collaboration process is simpler. The \texttt{Editor} attempts to fix the issue directly based on the code snippets retrieved by the \texttt{Searcher}.
Given the randomness in LLM code modifications, we draw on the approach similar to Agentless~\cite{xia2024agentless}, generating multiple candidate repair solutions in a single LLM request and normalizing the code using AST.
Finally, the model merges and votes on all candidate solutions, selecting the highest-voted one as the final repair solution.

\subsection{Code Indexing}

We have developed several code indexing tools with multilingual support, to satisfy different code search requirements for different software development tasks.

\subsubsection{Code Knowledge Graph}

A code knowledge graph represents code elements, their attributes, and the relationships between these elements in a graph structure, helping agents better understand and manage large codebases.
In this graph, vertices represent code entities (such as functions, variables, classes, etc.), and edges represent relationships between these entities (such as function calls, variable references, class inheritance, etc.).
This structured representation provides richer information about the codebase.

\toolname analyzes and organizes the code and documentation in a repository to generate a multi-directional graph using program analysis techniques.
This graph includes semantic nodes such as variables, functions, classes, and files, and edges representing file structure relationships, function call relationships, and symbol index relationships.
This results in a code knowledge graph that integrates code, documentation, and repository information from multiple data sources.

In the given codebase, each node and edge is uniquely identified, ensuring that every code entity is unique across the entire codebase.
The code knowledge graph uses graph attributes to store code entities and their dependencies.
Each node records its location, type, and name within the codebase, while each edge identifies the type of relationship between two nodes and the relationship's location in the code.

For example, consider the following code shown in Listing~\ref{lst:ckg_example}:

\begin{lstlisting}[language=Go, label={lst:ckg_example}, caption={Example of a code snippet in Go}]
// file: main/fileA.go
package main

import (
    "ckg-prompt/main/cmd/pkg_b"
    "fmt"
)

// StructA struct
type StructA struct{}

// FunctionA method for StructA
func (a *StructA) FunctionA() pkg_b.StructB {
    x := pkg_b.NewStructB() // Instantiate StructB
    return x
}

// XFunction function
func XFunction() {
    x := pkg_b.NewStructB() // Instantiate StructB
    x.FunctionB() // Calls FunctionB
}
\end{lstlisting}

\begin{figure}[h!]
    \centering
    \includegraphics[width=\linewidth]{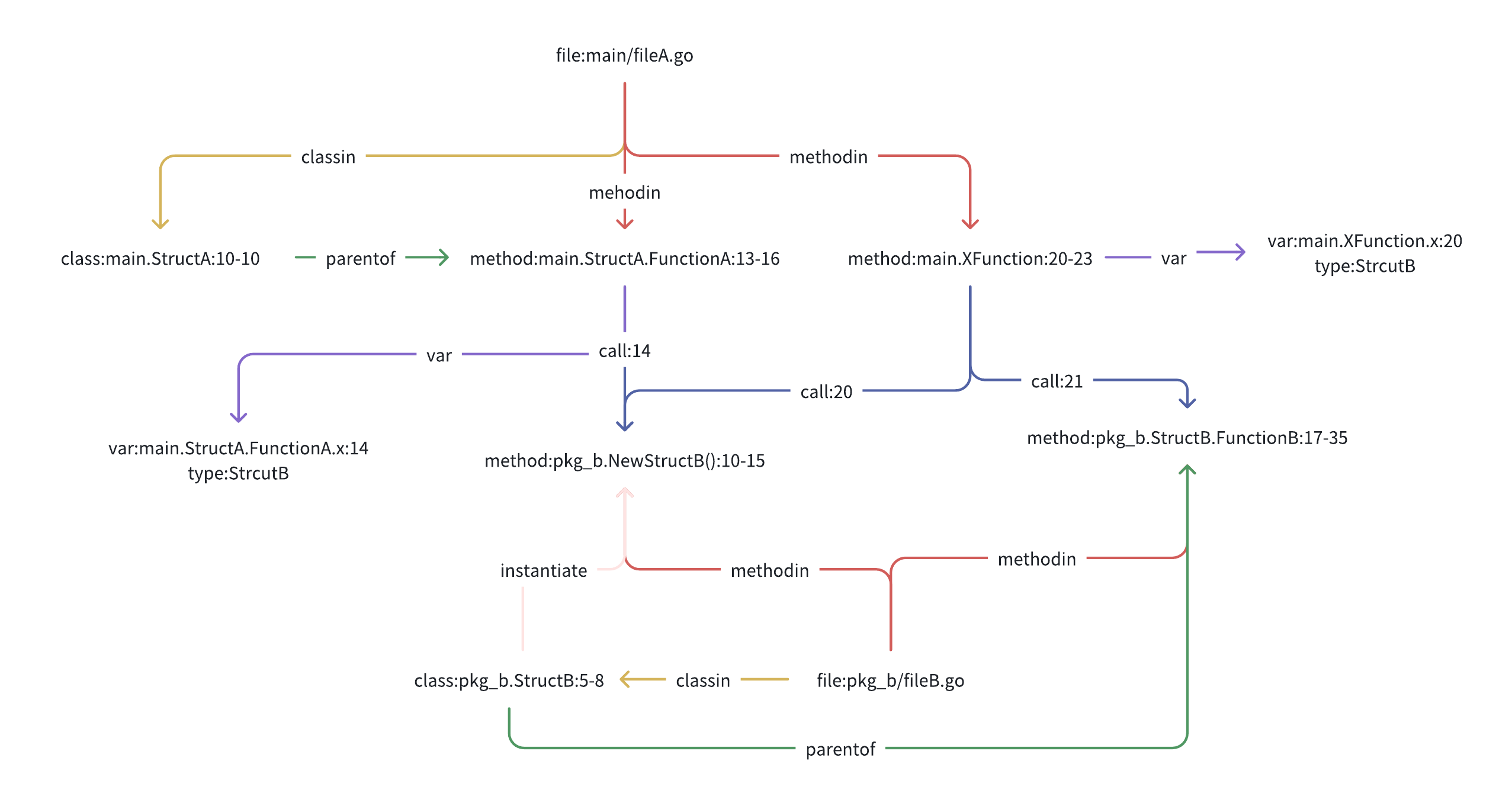}
    \caption{Code Knowledge Graph Example}
    \label{fig:ckg-example}
\end{figure}

The corresponding code knowledge graph is illustrated in Figure~\ref{fig:ckg-example}.

After constructing the code knowledge graph, the agent's code retrieval requests are processed through the following pipeline:

\begin{enumerate}
    \item The agent's query, along with any code statements, undergoes entity recognition using a model to identify entity mentions and types. These are then queried in the knowledge graph using SQL, resulting in candidate entity \texttt{list 1}.
    \item The agent's query, along with any code statements, is embedded and matched for similarity in the knowledge graph, yielding candidate entity \texttt{list 2}.
    \item The agent's query is directly converted into a search query through keyword recognition and queried in the knowledge graph using SQL, resulting in candidate entity \texttt{list 3}.
\end{enumerate}

The candidate entity \texttt{lists 1, 2, and 3} are then merged and ranked using a fine-ranking model to obtain the final entity \texttt{list X}, which is returned to the agent, completing the code retrieval process.

The code knowledge graph tool in \toolname enables comprehensive code retrieval, providing agents with repository-level knowledge question and answering (Q\&A) capabilities.
Currently, our code knowledge graph supports 12 common programming languages, including C, C\#, C++, Java, Kotlin, JavaScript, TypeScript, TSX, Rust, Go, Python, and Lua.

\subsubsection{Language Server Protocol}

The code knowledge graph can handle most class and function definition and reference retrieval needs in the target project, but it has the following limitations:

\begin{itemize}
    \item It cannot accurately retrieve definitions and references of classes, functions, and variables outside the target project (such as standard libraries and third-party libraries).
    \item For cases with multiple entities having the same name, the LSP can more accurately navigate to the relevant class or function definition, avoiding omissions or redundancies during the recall and re-ranking process.
\end{itemize}

\begin{figure}[h!]
    \centering
    \includegraphics[width=\linewidth]{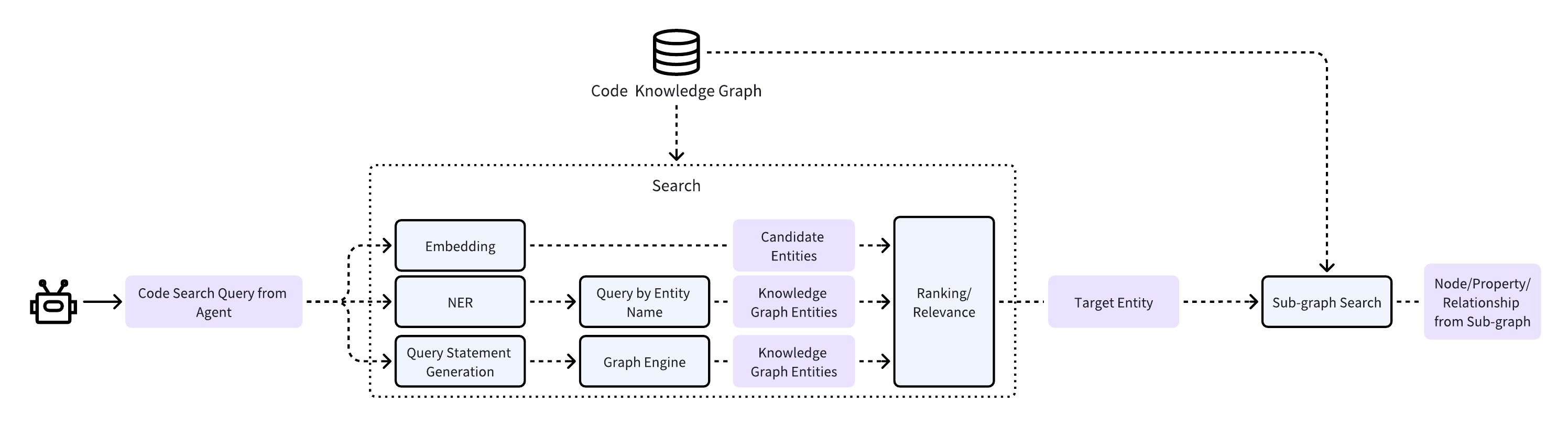}
    \caption{Code Retrieval using LSP}
    \label{fig:lsp}
\end{figure}

To address these issues, \toolname uses the language server protocol (LSP) to achieve global and precise code retrieval on the user's machine.
The LSP, developed by Microsoft, is widely compatible with various programming languages, markup languages, tools, and frameworks, making it highly versatile for IDE scenarios.
The process of code retrieval using the LSP in \toolname is illustrated in Figure~\ref{fig:lsp}.

The agent's use of the LSP for code retrieval is similar to a developer's \textit{Ctrl + Click} action in an IDE to jump to code.
However, since the agent's numerical positioning and computation capabilities are weak, we added fuzzy positioning features to enhance the agent's use of LSP tools:

\begin{itemize}
    \item Based on the file name and line number provided by the agent, search for identifiers in that line and compute the column number to form an LSP request.
    \item Based on the file name and line number provided by the agent, search for identifiers near that line and compute the column number to form an LSP request.
    \item Based on the identifier and line number provided by the agent, search for identifiers in the files the agent has opened and browsed to form an LSP request.
\end{itemize}

These services are prioritized from top to bottom, using the first successfully responded LSP request result as the tool's output.

\subsubsection{Other General Indexing Capabilities}

Besides LSP and the code knowledge graph, we also integrate general project file retrieval (find file), project or file identifier retrieval (grep), and other capabilities into the \toolname framework, providing a consistent toolset for code retrieval.

\subsection{Code Editing}

\subsubsection{Reflections on Code Editing}

In our long-term exploration of AI agents for software development, we tried various methods of using LLMs for code edit descriptions and found that current LLMs have generally weak code modification capabilities.
Below are some of the failed approaches we explored:

\begin{itemize}
    \item \textbf{Asking the agent to generate unified diff format code change descriptions.} The unified diff format presents the changes between the original file and the modified file in a unified manner, as examplified in Listing~\ref{listing:unified_diff}. The unified diff format has strict formatting requirements, and LLMs often struggle to correctly calculate line number increments, resulting in unapplicable unified diffs.
    \item \textbf{Asking the agent to provide the start and end line numbers and the replacement code snippet.} Even with line numbers added to all code retrieval results, LLMs, including GPT-4, often fail to provide the correct modification range, leading to issues like repeated lines or unintended deletions.
    \item \textbf{Rewriting the entire file.} Providing the entire file content and modification description to the LLM and asking it to output the modified file content avoids line number calculations but is economically unfeasible for each code edit and nearly unusable for long files. We are also working on obtaining a specialized code editing model through SFT for full-file rewriting, but this is a long-term plan.
\end{itemize}

\begin{lstlisting}[language=Go, label={listing:unified_diff}, caption={Example of Unified Diff}]
--- example.txt
+++ example.txt
@@ -1,4 +1,4 @@
 This is a sample file.
-It contains multiple lines of text.
-Here is another line.
-Goodbye!
+It contains a few lines of text.
+Here is yet another line.
+See you later!
\end{lstlisting}

Through extensive exploration and attempts, we concluded that LLM code edit descriptions need the following characteristics:

\begin{itemize}
    \item No strict format validation, with descriptions that can be stably applied after processing and parsing.
    \item No need to provide line number ranges or perform line number calculations, as LLMs are unstable in this aspect.
    \item Simple, concise descriptions to minimize token and time costs.
\end{itemize}

Inspired by Aider\footnote{\url{https://github.com/paul-gauthier/aider}}'s code change method, we developed our relatively stable code edit tool: \toolname \texttt{AutoDiff}.
\texttt{AutoDiff}'s code edit description resembles git conflict markers, where the agent provides the file path, original code, and replacement code within conflict markers.
\texttt{AutoDiff} parses the edit block, matches the provided original code snippet to the most similar segment in the file, and replaces it with the provided replacement code.
It then adjusts the indentation of the replacement code based on the modified file context.
Finally, the differences before and after the modification are compared to generate a unified diff format change file.
These modifications are simulated and not actually saved on the user's device, only to generate the unified diff format change file.
The final code modification requires subsequent static code diagnostics.

\subsubsection{Static Code Diagnostics}

\begin{figure}[htb!]
    \centering
    \includegraphics[width=\linewidth]{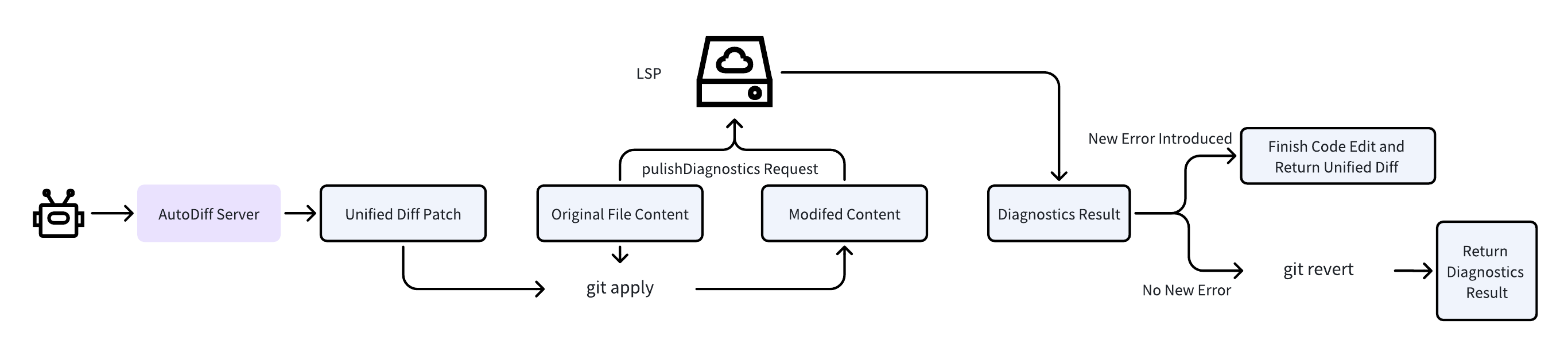}
    \caption{Static Code Diagnostics Workflow}
    \label{fig:diagnostics}
\end{figure}

Although \texttt{AutoDiff} can handle most code edit requests correctly, common syntax issues like type errors, undefined variables, indentation errors, and unclosed brackets still occur.
We use the LSP to perform static code diagnostics on files before and after \texttt{AutoDiff} modifications as shown in Figure~\ref{fig:diagnostics}

As shown in the figure, the workflow of static code diagnostics is as follows:

\begin{enumerate}
    \item Apply the \texttt{AutoDiff} generated unified diff format code edit patch to the original file to get the modified file content.
    \item Perform LSP static code diagnostics on the original file content and save the results.
    \item Perform LSP static code diagnostics on the modified file content and save the results.
    \item Compare the diagnostic results before and after the modification to check if new static errors (focusing on \textit{Fatal} and \textit{Error} levels) were introduced by the agent's modification.
    \item If no new errors were introduced, complete the modification and return a success message and the corresponding unified diff description to the agent.
    \item If new errors were introduced, return the relevant diagnostic information to the agent for further modifications and adjustments.
\end{enumerate}

\section{Experimental Results and Analysis}
We conducted a detailed evaluation of \toolname's performance on the SWE-bench Lite dataset.

\subsection{Dataset Overview: SWE-bench Lite}
SWE-bench, as introduced in Section~\ref{sec:swe-bench}, is a highly challenging benchmark for LLMs to solve program logic and functional bugs.
This dataset is consisted of 2294 issues from 12 industrial-grade Python code repositories on GitHub.
Given a codebase and a description of the issue to be resolved, the agent needs to retrieve and edit the code from the repository, ultimately submitting a code patch that resolves the issue.
Solving problems in SWE-bench typically requires understanding and coordinating changes across multiple functions, classes, or even files, necessitating interaction with the execution environment, handling extremely long contexts, and performing more complex reasoning than traditional code generation.
Evaluations in the SWE-bench paper show that directly applying Claude 2 and GPT-4 can only solve 4.8\% and 1.7\% of the instances, respectively~\cite{jimenez2023swe}.

Due to the high difficulty of SWE-bench, subsequent research found that evaluating on all 2294 instances of SWE-bench is a time and token-intensive process that is frustrating and does not validate short-term progress.
Therefore, the authors of SWE-bench extracted 300 instances with complete issue descriptions, clear-solving logic, and relative ease of resolution to form the SWE-bench Lite dataset.
Currently, the SWE-bench Lite dataset has become the benchmark for evaluating the capability of agents to solve software engineering problems, with over 20 companies and research organizations participating in the evaluation and submissions.

\subsection{\toolname Results}

In the latest SWE-bench Lite evaluation, \toolname successfully solved 102 instances, achieving a solve rate of 34\%. An analysis of this result is shown in Table~\ref{table:results}

\begin{table}[htb!]
\caption{Experimental Results on SWE-bench Lite}
\label{table:results}
\centering
\begin{tabular}{cc}
\hline
Item                                                 & Figure             \\ \hline
\# of Resolved Issues                                & 102                \\
Resolved Rate                                        & 102 / 300 = 34\%   \\
Precision of File Localization                       & 265 / 300 = 88.3\% \\
Precision Rate of Code Snippet Localization          & 206 / 300 = 68.7\% \\
Percentage of Issues Progressed to Dynamic Debugging & 84 / 300 = 28.0\%  \\
Percentage of Issues Progressed to Static Repair     & 202 / 300 = 72.0\% \\
Success Rate of Dynamic Debugging                    & 32 / 84 = 38.1\%   \\
Success Rate of Static Repair                        & 70 / 216 = 32.4\%  \\ \hline
\end{tabular}
\end{table}

We compared the files containing code snippets located by the agent during the solving process with the files containing the gold patches for the instances.
If the file containing the gold patch was included, it was considered a successful file localization.
Similarly, if there was an inclusion or overlap relationship between the code snippets found during the solving process and the modification target segments of the gold patch, the solving process was considered to have successfully located the target code segment.

\begin{figure}[htb!]
    \centering
    \includegraphics[width=0.7\linewidth]{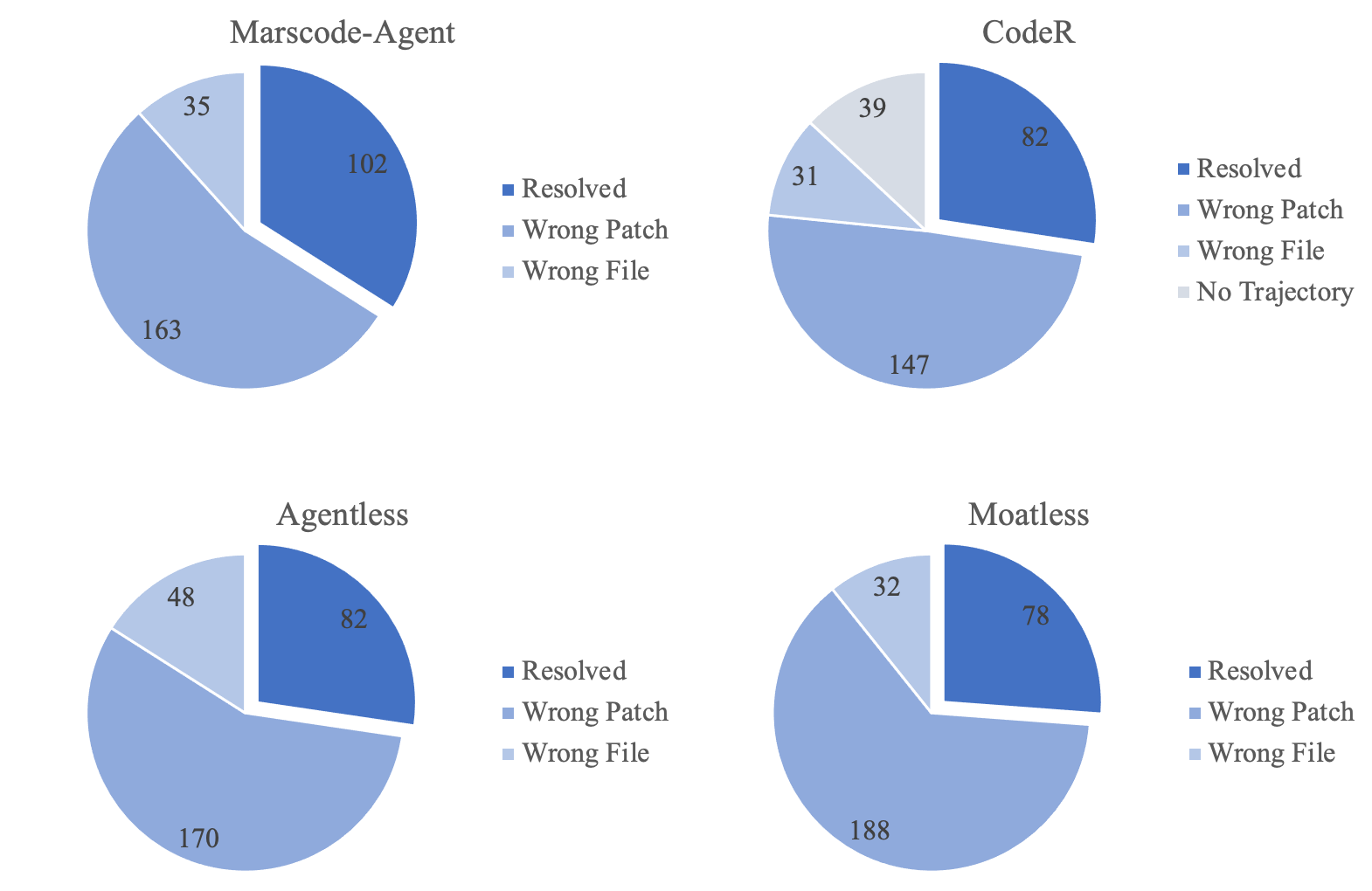}
    \caption{Correct Patch and File Localization Comparison among Tools}
    \label{fig:comparison}
\end{figure}

\begin{figure}[htb!]
    \centering
    \includegraphics[width=0.7\linewidth]{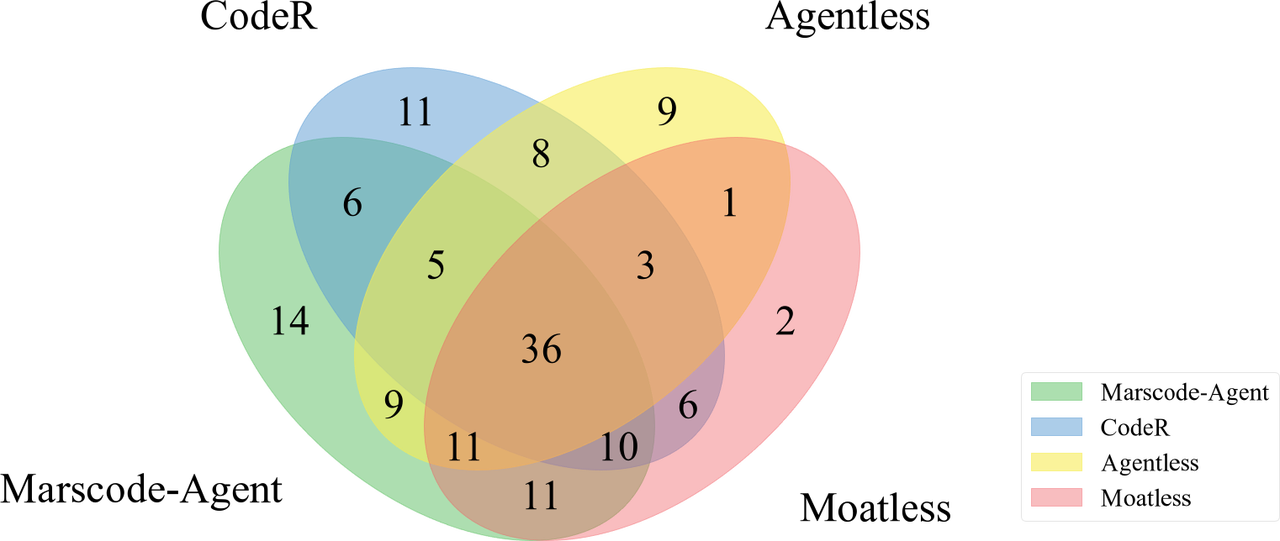}
    \caption{Success Rate Comparison among Tools}
    \label{fig:comparison2}
\end{figure}

Using the same analysis method, we compared the code retrieval effectiveness of currently publicly available traceable solutions (CodeR~\cite{chen2024coder}, Moatless\footnote{\url{https://github.com/aorwall/moatless-tools}}, Agentless~\cite{xia2024agentless}) as shown in Figures~\ref{fig:comparison} and~\ref{fig:comparison2}.
\toolname, using code knowledge graphs, language server procotols, and other code retrieval tools, successfully located the files to be modified in 265 out of 300 instances (88.3\% success rate, higher than the current leader Aide's 78\% file localization accuracy with a solve rate of 43\%) and successfully located the target segments in 206 instances. From the perspective of code retrieval and error localization capabilities, \toolname is in a leading position.

\begin{figure}[htb!]
    \centering
    \includegraphics[width=0.5\linewidth]{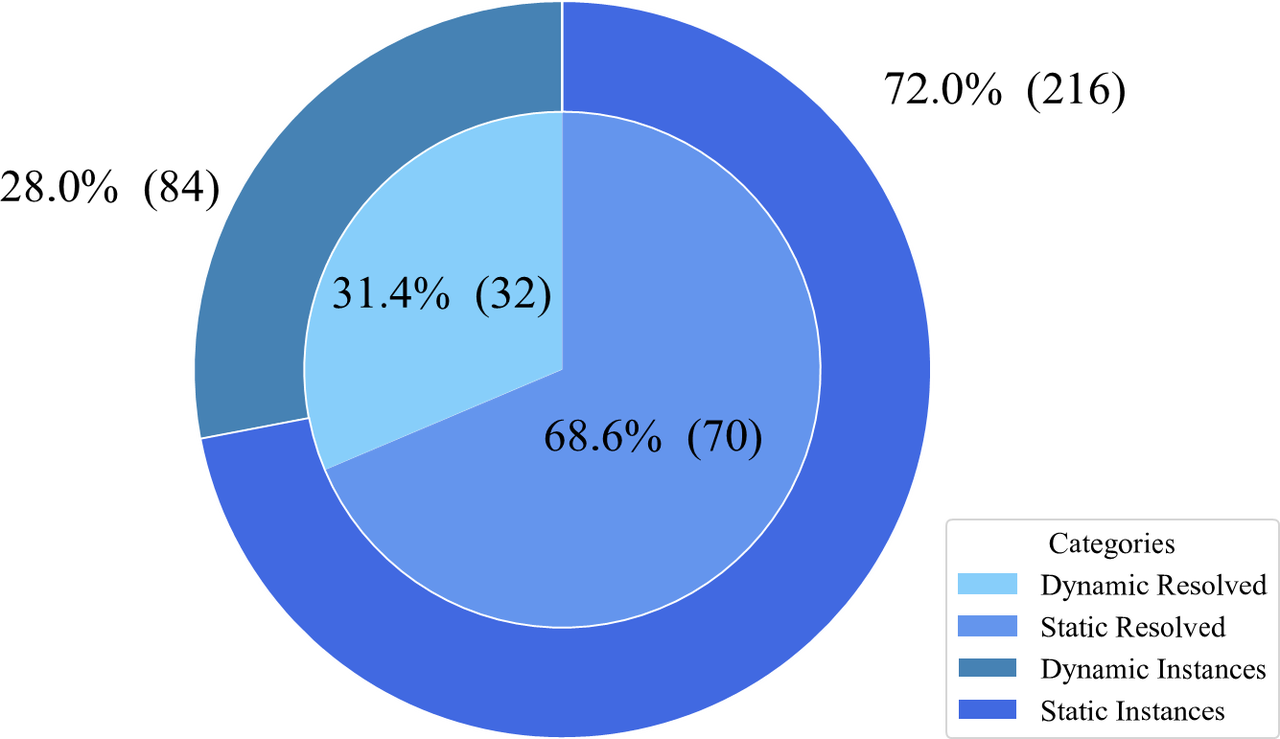}
    \caption{Distribution of Dynamic and Static Traces by \toolname}
    \label{fig:dynamic_and_static}
\end{figure}

We analyzed the distribution of instances solved by static and dynamic methods in the experiment, as shown in Figure~\ref{fig:dynamic_and_static}.
Of all instances, $84/300=28\%$ were considered suitable for dynamic solving by the Planner Agent, while $216/300=72\%$ were considered suitable for static solving.
Dynamic solving successfully resolved 32 instances, with a solve rate of 38.1\%, while static solving successfully resolved 70 instances, with a solve rate of 32.4\%.
Due to the process of defect reproduction and verification in dynamic debugging, the solve rate is slightly higher than that of static repair.

\section{Final Remarks}
In this paper, we introduced \toolname, a novel framework leveraging LLMs to automate bug fixing and software development tasks.
Our approach combines advanced code analysis techniques with LLM capabilities to provide a systematic process for fault localization, candidate patch generation, and patch validation.
Through comprehensive evaluations on the SWE-bench Lite dataset, \toolname demonstrated significant improvements in solving real-world software engineering problems, achieving a solve rate of 34\%.

Looking forward, we aim to further enhance \toolname by reducing LLM call costs, improving user-agent collaboration, supporting dynamic debugging within ther real user workspaces to avoid environmental contamination, and increasing the accuracy of error localization and code modifications.
Our ongoing commitment is to refine and expand the capabilities of \toolname, making it an indispensable tool in the landscape of intelligent software development.

\toolname's promising results on SWE-bench Lite demonstrates the potential of LLMs to significantly advance the field.
We hope our work inspires further research and development, driving innovations that bring us closer to fully autonomous software engineering solutions.

\bibliography{references}{}
\bibliographystyle{plain}

\end{document}